\documentclass[a4paper,11pt]{article}
\usepackage{pos}
\usepackage[caption=false]{subfig}

\newcommand{\GEANT}{{\textsc{Geant}}}

\title{Latest oscillation results from Daya Bay}
\author*[a]{Jinjing Li}

\affiliation[a]{Tsinghua University,\\
No.~30, Shuangqing Road, Haidian District, Beĳing, PRC}

\emailAdd{jinjing-li@mail.tsinghua.edu.cn}
\note{For the Daya Bay collaboration.}

\abstract{
    The Daya Bay reactor neutrino experiment, which pioneered the measurement of a non-zero value for the neutrino mixing angle $\theta_{13}$ in 2012, operated for nearly nine years, from Nov.~24, 2011, to Dec.~12, 2020. Antineutrinos produced by six reactors, each with a thermal power of 2.9 GW, were detected by eight identically designed detectors located in two near and one far underground experimental halls. The experimental setup, featuring kilometer-scale baselines between the detectors and reactors, enabled precise investigations within the three-neutrino mixing framework. This proceeding presents the latest measurements of the neutrino mixing angle $\theta_{13}$ and the mass-squared difference $\Delta m_{32}^2$, based on samples tagged via gadolinium capture ($n$Gd) and hydrogen capture ($n$H). The combined results from Daya Bay's most recent $n$Gd and $n$H data sets provide a $2.6\%$ precision for $\sin^22\theta_{13}$. Additionally, a search for light sterile neutrinos with full data set is summarized.
    }

\FullConference{42nd International Conference on High Energy Physics (ICHEP2024)\\
18-24 July 2024\\
Prague, Czech Republic\\}

\begin{document}
\maketitle

\section{The Daya Bay Reactor Neutrino Experiment}
The Daya Bay reactor neutrino experiment, located near Shenzhen, China, was designed to precisely measure the neutrino mixing angle $\theta_{13}$ by studying electron antineutrinos ($\overline{\nu}_e$) emitted from six nuclear reactors. Each reactor produced a thermal output of up to 2.9 GW. The experimental setup consisted of eight identically designed antineutrino detectors (ADs) distributed across three underground experimental halls (EHs). EH1 and EH2, each housing two ADs, were located approximately 500 meters from the reactor cores, enabling precise measurements of the $\overline{\nu}_e$ flux and energy spectrum. EH3, situated roughly 1.6 kilometers from the reactors, contained four ADs and was positioned to observe the first oscillation minimum of $\overline{\nu}_e$ modulated by $\sin^2 2\theta_{13}$, as described by the following equation:
\begin{equation}
    \begin{aligned}\label{eq:oscillation_prob}
        P_{\overline{\nu}_e \rightarrow \overline{\nu}_e} & = 1 - \cos^4 \theta_{13} \sin^2 2\theta_{12} \sin^2 \Delta_{21} - \sin^2 2\theta_{13} \left( \cos^2 \theta_{12} \sin^2 \Delta_{31} + \sin^2 \theta_{12} \sin^2 \Delta_{32} \right) \\
                                                          & = 1 - \cos^4 \theta_{13} \sin^2 2\theta_{12} \sin^2 \Delta_{21} - \sin^2 2\theta_{13} \sin^2 \Delta_{\mathrm{ee}},
    \end{aligned}
\end{equation}
where $\Delta_{ij} = 1.267 \Delta m_{ij}^2 \left(\text{eV}^2\right) \left[L~(\text{m}) / E~(\text{MeV})\right]$ represents the phase difference due to the mass-squared splitting between neutrino states $i$ and $j$, $L$ is the baseline distance, and $E$ is the neutrino energy. The effective mass-squared difference $\Delta m^2_{\mathrm{ee}}$ governs the oscillation wavelength observed in the Daya Bay experiment.

The multi-baseline strategy mitigated reactor-related systematic uncertainties, such as variations in $\overline{\nu}_e$ flux and fuel composition, by allowing relative comparisons between near and far detectors. The $\overline{\nu}_e$ were detected via the inverse beta decay (IBD) process: $\overline{\nu}_e + p \rightarrow e^+ + n$. The liquid scintillator (LS), which contained protons for the IBD interaction, served as the primary medium for $\overline{\nu}_e$ detection. The central target for oscillation studies focused on neutron capture on gadolinium ($n$Gd), using 20 tonnes of LS doped with $0.1\%$ gadolinium by weight (GdLS), housed in a 3-meter-diameter inner acrylic vessel (IAV). Surrounding this was a 4-meter-diameter outer acrylic vessel (OAV), filled with undoped LS, acting as a gamma catcher and later as the primary target for neutron capture on hydrogen ($n$H) in oscillation analyses. Further details on the experimental configuration and detector design are available in Ref.~\cite{DayaBay:2015kir}.

The positron and neutron produced from the IBD reaction created a distinct time-correlated signal pair. The positron deposited its energy promptly in the scintillator, while the neutron was predominantly captured by gadolinium or hydrogen, producing a delayed signal. Data collection spanned nearly nine years, from December 24, 2011, to December 12, 2020. Detailed descriptions of event selection, calibration, and reconstruction methods can be found in Ref.~\cite{DayaBay:2016ggj}.

\section{Neutrino Oscillation Results Based on Neutron Capture on Gadolinium}
The latest $n$Gd analysis from Daya Bay utilized $5.55 \times 10^6$ $n$Gd IBD candidates collected over 3158 days~\cite{DayaBay:2022orm}. This analysis features various improvements in IBD selection, energy calibration, and background treatment compared to previous analysis~\cite{DayaBay:2018yms}. Corrections for energy response non-uniformity and non-linearity were applied using spallation neutron (SPN) data and delayed $\alpha$ particles from $^{214}\mathrm{Bi}-^{214}\mathrm{Po}$'s cascade decay. A new non-linearity correction curve, derived from flash-ADC data collected in 2016, was used to calibrate the electronics' response for each channel.

New photomultiplier tube (PMT) flashers were identified and removed by refining selection criteria based on charge distributions and timing patterns, successfully eliminating over $99\%$ of such events with negligible inefficiency for IBD signals. The dominant background remains cosmogenic $^9\mathrm{Li}/^8\mathrm{He}$. A new technique was developed to classify the  $^9\mathrm{Li}/^8\mathrm{He}$-enhanced sample based on muon energy, and the spatial and temporal separations between prompt and delayed signals, allowing for a more precise determination of the $^9\mathrm{Li}/^8\mathrm{He}$ spectrum and reducing uncertainty to below $25\%$.

Additionally, a newly identified muon-induced background, referred to as ``muon-x'', arose due to PMT failures. This background was mitigated by rejecting events with a delayed signal occurring within 410 $\mu$s of the specific muon event, with minimal loss of live time. The background rate was determined by extending the prompt energy spectrum up to 250 MeV and fitting for fast neutron and muon-x components.

\begin{figure}[!htb]
    \centering
    \subfloat[]{\includegraphics[width=7cm]{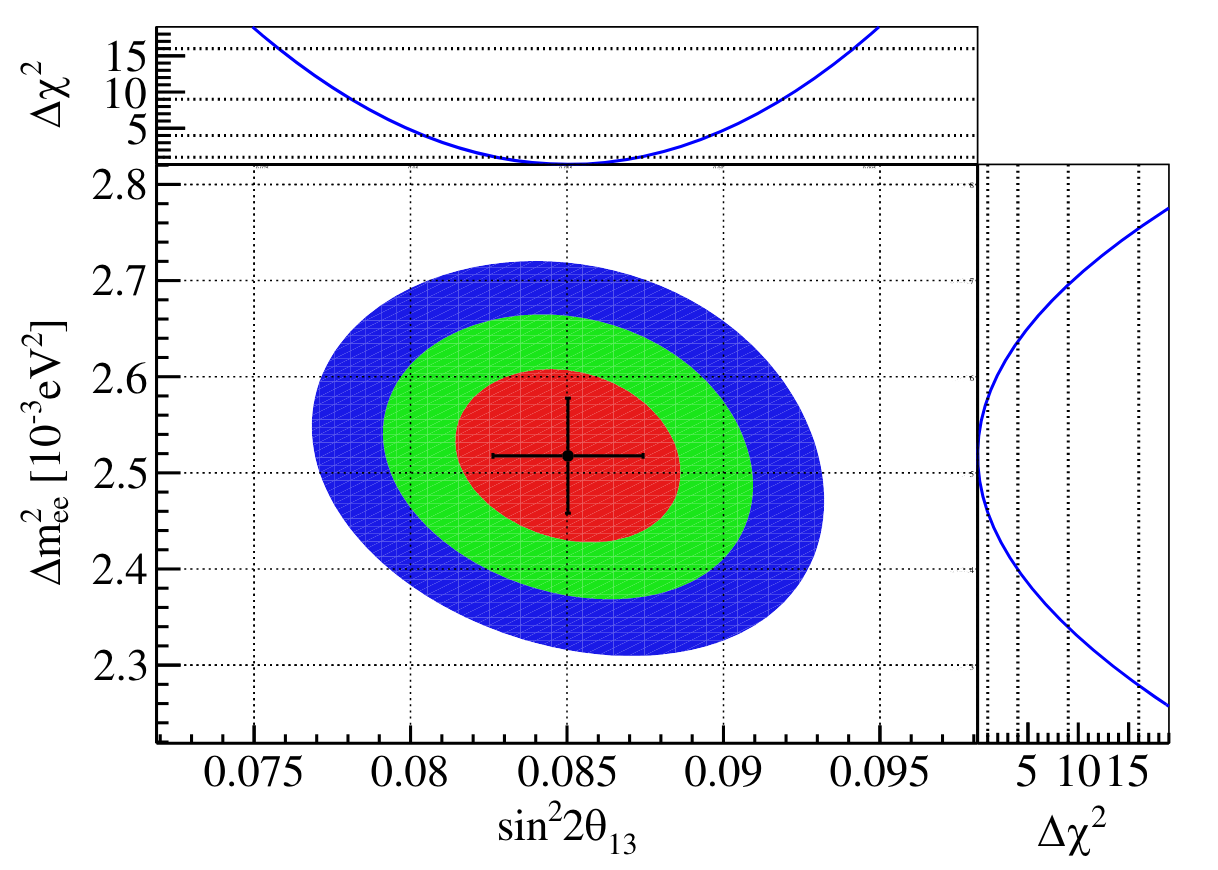}}
    \subfloat[]{\includegraphics[width=7cm]{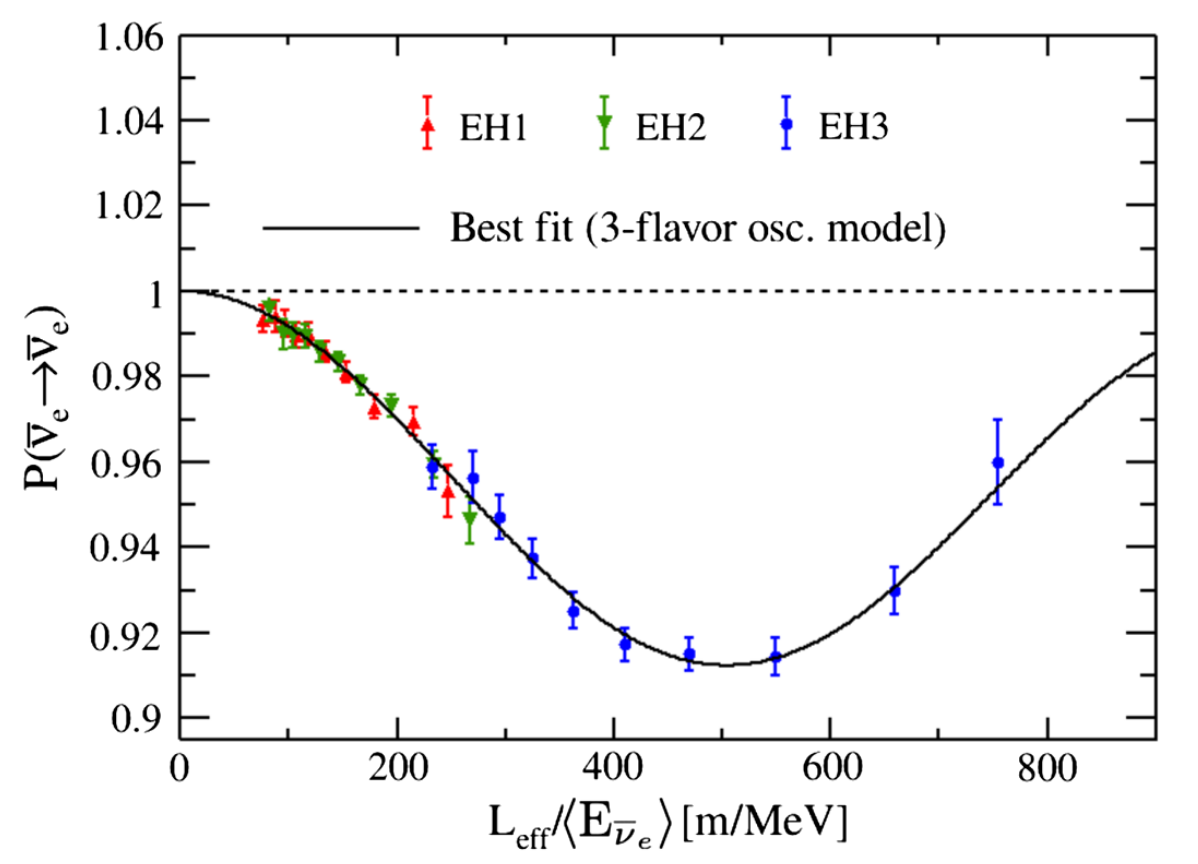}}
    \caption{(a) Contours and best-fit point in the $\Delta m^2_{\mathrm{ee}}$ vs.~$\sin^2 2\theta_{13}$ plane. (b) Measured survival probability vs.~$L_{\mathrm{eff}}/\langle E_{\overline{\nu}_e} \rangle$.}
    \label{fig:nGd_fitting_contour}
\end{figure}

Oscillation parameters were extracted following the method outlined in Ref.~\cite{DayaBay:2016ggj}. The best-fit results are $\sin^2 2\theta_{13} = 0.0851 \pm 0.0024$, with $\Delta m^2_{32} = (2.466 \pm 0.060) \times 10^{-3}~\mathrm{eV}^2$ for normal mass ordering (NO), and $-(2.571 \pm 0.060) \times 10^{-3}~\mathrm{eV}^2$ for inverted mass ordering (IO). Figure~\ref{fig:nGd_fitting_contour} illustrates the fit contours and the measured survival probability across the three halls as a function of $L_{\mathrm{eff}}/\langle E_{\overline{\nu}_e} \rangle$.

\section{Neutrino Oscillation Results Based on Neutron Capture on Hydrogen}
Daya Bay recently presented the first analysis of $\overline{\nu}_e$ oscillation amplitude and frequency using the $n$H channel, based on 1958 days of data~\cite{DayaBay:2024hrv}. We utilized over 3.6 million IBD candidates, marking a significant increase in statistics compared to the previous study~\cite{DayaBay:2016ziq}. It incorporates various improvements in event selection, background estimation, and systematic uncertainty, etc.

Two independent analyses, A and B, were performed, yielding consistent results. The result of Analysis A which achieves a better precision is chosen as nominal. This proceeding focuses solely on Analysis A. The analysis builds on the reconstruction outlined in Ref.~\cite{DayaBay:2016ggj}, primarily using SPN events tagged by $n$Gd to calibrate the energy response, optimized for the GdLS region. The energy response non-uniformity was further calibrated using the $2.2$-MeV $\gamma$ from $n$H SPNs, ensuring that the energy scales for $n$H IBDs were aligned. After calibration, the energy scale difference among the ADs was controlled to $<0.30\%$, with residual non-uniformity within each AD kept below $0.5\%$.

A new combined cut, $DT\equiv \text{distance}+\text{time}\times v < 1~\mathrm{m}$, where $v=1~\rm{m}/0.6~\rm{ms}$ (approximating the speed of thermalized neutrons), was proposed to select $n$H IBDs. This cut improved the signal-to-background ratio and reduced the uncertainty in the relative efficiency. The optimal $DT$ cut value maximized sensitivity to $\theta_{13}$, and the results were shown to be robust against variations in the cut.

Accidental background rates were calculated using measured single rates ($R_s$) following the methods described in Ref.~\cite{Yu:2013cob}. Compared to earlier studies, $R_s$ were extracted through iterations of the prediction equation, resulting in negligible systematic uncertainty. The uncertainties in accidental rates and spectra were minimized to a negligible level ($0.03\%$) after validation using data-driven techniques. The fast neutron, Am-C, and $^9$Li/$^8$He backgrounds were reevaluated using the established methodology from Ref.~\cite{DayaBay:2016ziq}.

A new correlated background from neutrons produced in the spontaneous fission of $^{238}$U or via $(\alpha,n)$ reactions in PMT glass was evaluated. Neutron recoils or $\gamma$ rays from fission and deexcitation processes can mimic the prompt signal. The content of boron trioxide, which provides main target of $(\alpha,n)$, i.e., $^{11}$B, were measured in five PMT glass samples of known weight. A \GEANT4-based simulation, utilizing fission data from FREYA~\cite{verbeke2018fission} and ($\alpha,n$) reaction cross-sections from the JENDL/AN-2005 database~\cite{murata2006evaluation}, was used to evaluate the background. This new radiogenic neutron background was estimated at $0.20\pm0.04/\text{day}/\text{AD}$, impacting the rate-only fit but having minimal effect on the rate+shape fit.

The uncertainty in the 1.5 MeV prompt-energy cut efficiency was induced by a $0.3\%$ relative variation in energy scale and differences in energy leakage between ADs, evaluated at $0.08\%$ and $0.10\%$, respectively, yielding a total uncertainty of $0.13\%$.

The delayed-energy cut efficiency uncertainty was assessed using the method introduced in Ref.~\cite{DayaBay:2016ziq}. We first compared the number of IBD candidates selected within three times the Gaussian width of the $n$H peak to those in the broader $[1.5,~2.8]$~MeV range. The AD-to-AD differences in the $n$H SPN-to-$n$Gd SPN ratio, sensitive to differences in capture ratios and energy leakage, were also evaluated, yielding a consistent systematic uncertainty of $0.20\%$.

For the $DT$ cut, a new IBD sample was generated by relaxing the cut to 3~m and subtracting the accidental background, which served as the denominator for estimating the $DT$ cut efficiency in each AD. The $1\sigma$ spread of the resulting efficiencies for the eight ADs was used to quantify the uncertainty. This procedure was repeated on samples from $^{214}$Bi-$^{214}$Po-$^{210}$Pb cascade decays and $n$H IBDs with a higher prompt-energy cut ($E_p>3$ MeV), yielding consistent uncertainty at $0.20\%$.

The total systematic variation in the number of detected $\overline{\nu}_e$ events among detectors was estimated at $0.34\%$. Correlated efficiency uncertainties between ADs largely cancel out due to the relative near-far measurement and are therefore not relevant. Some uncertainties were checked and remain consistent with previous results, and thus are not discussed here.

We developed an energy response model to establish the relationship between the actual $\overline{\nu}_e$ energy and the reconstructed prompt energy. One of the key residual uncertainties stems from the energy leakage differences due to the IAV and OAV geometrical differences among the ADs. The model also addresses the nonlinear transformation of deposited energy into scintillation light, which is then detected by the PMTs and their associated electronics. Data-driven corrections are applied to mitigate the impact of non-uniform energy scale and resolution effects. Furthermore, the model corrects for small distortions in the prompt energy spectrum induced by the $DT$ cut. This correction is derived by comparing the energy spectrum obtained under the nominal $DT$ cut with a loosened 3~m $DT$ cut.

\begin{figure}[!h]
    \centering
    \subfloat[]{
        \includegraphics[width=0.44\columnwidth]{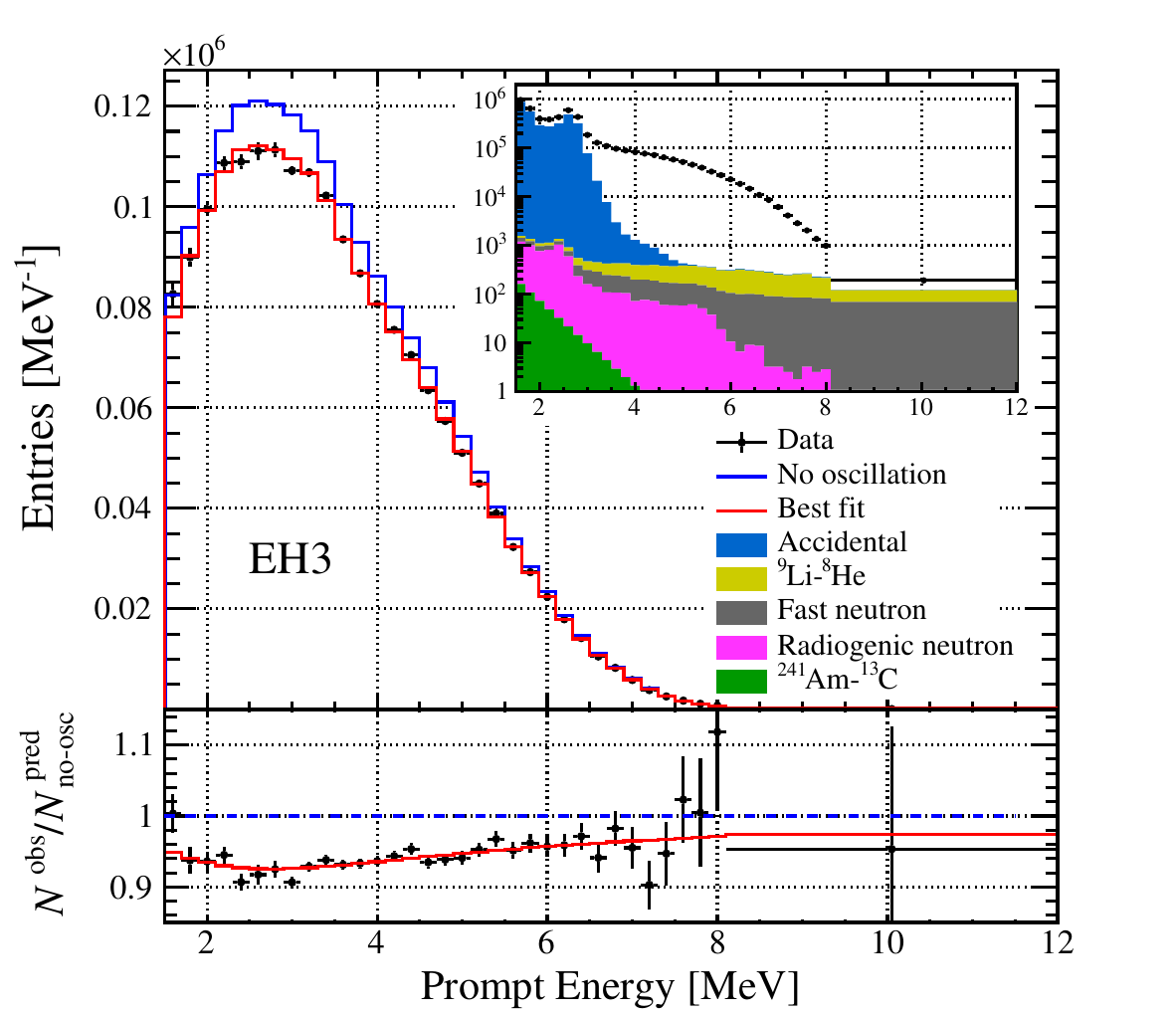}\label{fig:MvsP}
    }\hspace*{\fill}
    \subfloat[]{
        \includegraphics[width=0.50\columnwidth]{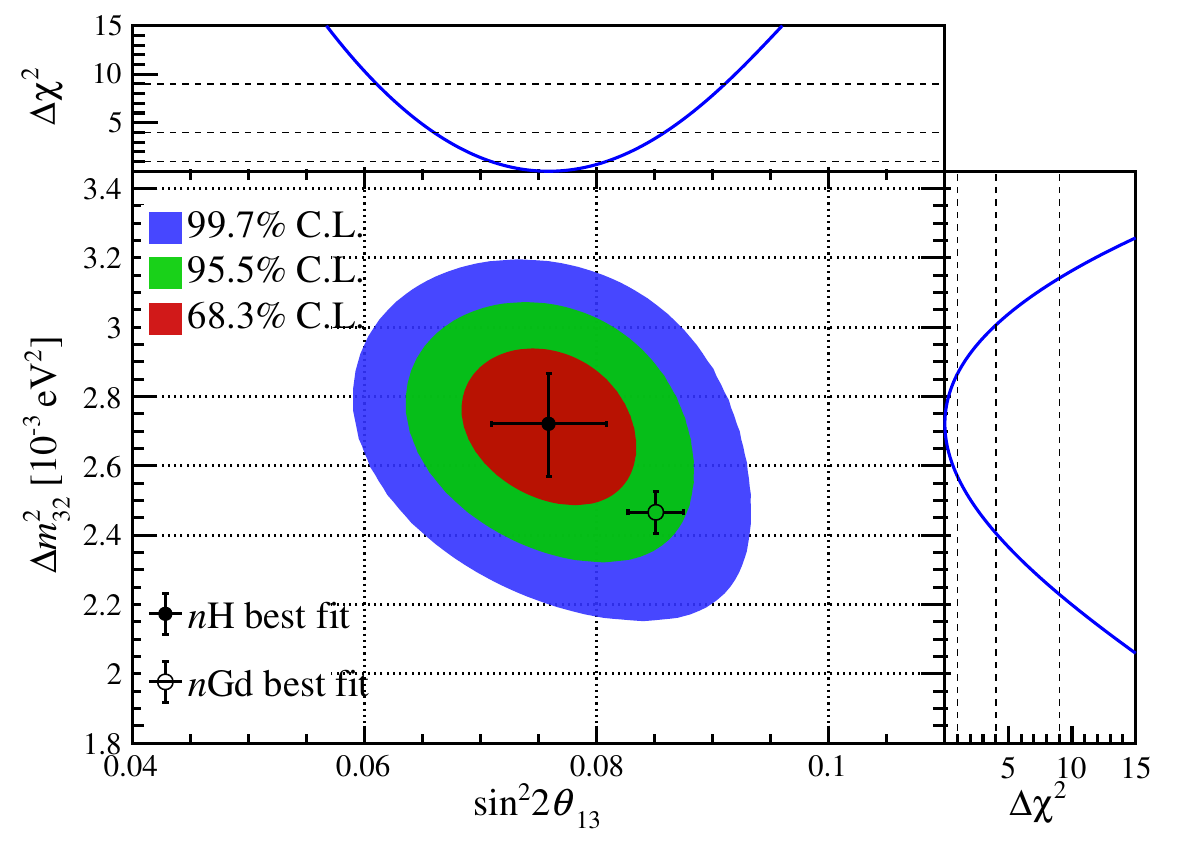}\label{fig:contour}
    }
    \caption{(a) Measured and best-fit spectrum distortion at EH3; (b) Confidence regions and best-fit points in the $\Delta m_{32}^2$ vs.~$\sin^22\theta_{13}$ plane.}\label{fig:nH}
\end{figure}

Oscillation parameters were determined through a $\chi^2$ fit, incorporating the measured $\overline{\nu}_e$ rate and spectrum from each AD, alongside the survival probability defined in Eq.~\eqref{eq:oscillation_prob}. The fit considers the correlated variations in reactor $\overline{\nu}_e$ rates and spectra to address potential model discrepancies. The best-fit results are: $\sin^2 2\theta_{13} = 0.0759^{+0.0050}_{-0.0049}$, $\Delta m^2_{32} = (2.72^{+0.14}_{-0.15}) \times 10^{-3} \, \text{eV}^2$ (NO), and $\Delta m^2_{32} = (-2.83^{+0.15}_{-0.14}) \times 10^{-3} \, \text{eV}^2$ (IO), with a $\chi^2/\text{NDF} = 256.7/236$. Figure~\ref{fig:nH}~\subref{fig:MvsP} compares the observed and predicted prompt energy spectra in the far hall, both with and without oscillation, using the best-fit parameters. Figure~\ref{fig:nH}~\subref{fig:contour} illustrates the best-fit result and the confidence regions for $\Delta m^2_{32}$ and $\sin^22\theta_{13}$ at 68.3\%, 95.5\%, and 99.7\% confidence levels.

These results are consistent with Daya Bay's $n$Gd measurements, offering high-precision independent validation of the oscillation parameters. Since there is negligible correlation between the $n$H and $n$Gd results, combining the two datasets enhances precision. The combined result gives a final value for the mixing angle of $\sin^2 2\theta_{13} = 0.0833 \pm 0.0022$, marking an 8\% improvement in precision over the $n$Gd-only result, making this the most precise measurements of $\theta_{13}$ to date.

\section{Search for Light Sterile Neutrino}
Neutrino oscillation phenomena, as observed across numerous experiments, are largely consistent with the three-neutrino mixing framework. Nonetheless, anomalous results, such as the excess of electron-like events reported by the LSND and MiniBooNE experiments in $\nu_\mu$/$\overline{\nu}_\mu$ beams at short baselines, as well as the reactor antineutrino anomaly observed in short-baseline reactor studies, have raised the possibility of additional, light neutrino states beyond the three known active ones.

The Daya Bay experiment conducted a comprehensive search for sterile neutrino mixing in the sub-eV range, analyzing its full dataset of approximately 5.55 million inverse beta decay (IBD) events collected over 3158 days of operation~\cite{DayaBay:2024nip}. This investigation was performed within the 3+1 neutrino mixing model, where the existence of a sterile neutrino is hypothesized to mix with the standard active neutrinos. Thanks to the expanded dataset and significant reductions in systematic uncertainties, this study achieved greater sensitivity than prior efforts.

No evidence of sterile neutrino oscillations was observed. Exclusion limits on the mixing parameters were determined using both the Feldman-Cousins and CLs statistical approaches. The analysis attained world-leading sensitivity, constraining $\sin^2 2\theta_{14}$ to values as low as $5 \times 10^{-3}$ at the 95\% confidence level, covering the mass-squared difference range $2 \times 10^{-4} \, \text{eV}^2 \lesssim \Delta m^2_{41} \lesssim 2 \times 10^{-1} \, \text{eV}^2$. These exclusion contours provide world-leading constraints on sterile-active neutrino mixing in this mass-splitting range, significantly improving upon previous reactor neutrino experiment limits. Notably, sterile neutrino mixing with $\sin^2 2\theta_{14} \gtrsim 0.01$ is excluded at the 95\% confidence level for $\Delta m^2_{41}$ between $0.01 \, \text{eV}^2$ and $0.1 \, \text{eV}^2$.

\section{Summary}
Daya Bay collaboration has reported the most precise measurement result of $\theta_{13}$ to date and one of the best measurements of $\Delta m^2_{32}$, using the complete data set with 3158 days of operation. Recently, Daya Bay has published the first analysis of oscillation amplitude and frequency using the independent $n$H IBD sample with 1958 days of data, giving a high-precision measurement result for $\theta_{13}$ and $\Delta m_{32}^2$. A combination of Daya Bay's latest result $\theta_{13}$ results from $n$Gd and $nH$ analyses is also reported, present a $2.6\%$ precision of $\sin^22\theta_{13}$, which is a $8\%$ improvement to $n$Gd only result. Using Daya Bay's complete data sample, no visible oscillation due to mixing of a sub-eV sterile neutrino with active neutrinos was found. World-leading exclusion limits are set by both Feldman-Cousins and CLs methods in $\Delta m_{41}^2$ vs.~$\sin^22\theta_{14}$ space.
\bibliographystyle{apsrev4-1}
%\begin{thebibliography}{99}
\bibliography{ref}
%\end{thebibliography}

\end{document}